# Rotational migration in human pancreatic ductal organoids depends on actin and myosin activity.

Gengqiang Xie[1#], Chaity Modak[1#], Olalekan H Usman[1], Raphael WF Tan[1], Nicole Coca[1], Gabriela De Jesus[1], Yue Julia Wang[1], D. Thirumalai[2], Xin Li[2*], Jerome Irianto[1*]

[1]Department of Biomedical Sciences, College of Medicine, Florida State University, Tallahassee, FL 32306

[2]Department of Chemistry, University of Texas at Austin, Austin, TX 78712; Department of Physics, University of Texas at Austin, Austin, TX 78712

[#]co-first authors

*co-corresponding authors:

Jerome Irianto, Email: jerome.irianto@med.fsu.edu, Xin Li, Email: xinlee0@gmail.com

**Running title:** Rotational migration of pancreatic ductal organoid

**Abstract**

Rotational migration is one specific form of collective cell migration when epithelial cells are confined in a spherical geometry, such as in the epithelial acini. This tissue-level rotation motion is crucial for the morphogenesis of multiple epithelial systems. Here, we introduce a new primary human model for the study of rotational migration, pancreatic ductal organoids. Live imaging revealed the persistent rotation of the organoids over time. By tracking the nuclei, the three-dimensional trajectory of the cellular movement was reconstructed and the velocity of the rotation was quantified. The presence of focal adhesion clusters and prominent actin stress fibers were observed at the basal side of the organoids, suggesting the interactions between the cells and the surrounding extracellular matrix. Finally, our inhibition study showed the dependence of pancreatic ductal organoid rotational migration on myosin activity, actin polymerization, and actin branching. We hope that this model will enable future studies with human primary cells, which are more faithful to normal epithelial cells.

**Introduction**

The collective migration of epithelial cells plays a vital role in many biological processes, including morphogenesis, wound repair, tissue homeostasis, and cancer metastasis[1]. One specific form of collective migration is rotational migration, which is observed when epithelial cells are confined in a spherical geometry. In rotational migration, the entire multicellular structure rotates as a result of directed collective migration. This migration mode has been observed in various systems, such as the follicle cells of *Drosophila* egg chambers[2-6], canine kidney epithelial acini[7-9], murine pancreatic organoids[10,11], and human mammary acini[12-14]. This suggests that rotational migration is an inherent dynamic feature of three-dimensional multicellular structures. Moreover, the significant role of rotational migration in morphogenesis has been reported in several studies. For example, disruption of *Drosophila* follicle cells' rotational migration prevented the elongation of the egg chamber[15,16], a critical step in its developmental process. Recent work in human mammary gland morphogenesis has also revealed that rotational migration is needed for the budding of spherical alveoli from the mammary duct network[17]. Furthermore, the loss of rotational migration was associated with the malignant behavior of breast cancer cells[12,13], suggesting the potential role of rotational migration in maintaining normal tissue homeostasis.

To establish the collective rotational migration, the epithelial cells need to collectively polarize towards the migration direction. Interestingly, unlike other collective cell migration, polarizing external cues such as space or chemical gradient are not available in rotational migration. Hence, the polarization process is likely to rely on cell-to-cell or cell-to-extracellular matrix (ECM) interactions. Several studies using the invertebrate *Drosophila* follicle cell model have identified a number of these polarizing cell-cell interactions. The accumulation of cadherin protein Fat2 at the trailing edge of a follicle cell is essential for the polarization of the WAVE complex at the leading edge of the adjacent cell[2]. In another study, they showed the roles of formin protein DAAM and Rho GTPase Cdc42 in actin stress fibers and protrusions polarization[3,4]. For the mammalian systems, on the other hand, the cellular polarization mechanisms are still unknown. Nevertheless, perturbation of some key players, such as actomyosin contractility[8,12,13,18], cell-cell junction protein E-cadherin[18], ECM receptor integrin β1[12,13], signaling proteins EGFR[12], and β-catenin[9], has been shown to disrupt the rotational migration in mammalian models. However, most of the findings were established in immortalized cell lines. In our opinion, the study of rotational migration regulation will benefit from primary human models, derived from human tissues and more faithful to normal epithelial cells. This will potentially provide more accurate findings on epithelial cell behavior and biology. Recent developments in organoid culture systems open the door for *in vitro* primary human models, such as in the brain[19], intestine[20], pancreas[21], and others[22,23].

In this study, we introduce a new human model for rotational migration study, the pancreatic ductal organoids, which were derived from human exocrine tissues. Upon culture in Matrigel, the pancreatic ductal epithelial cells form a cystic structure consisting of a spherical lumen encapsulated by a single layer of epithelial cells. Live imaging revealed the persistent rotation of the organoids over time. By tracking the nuclei, we reconstructed the trajectory of the cellular movement and quantified its velocity. We also observed the presence of focal adhesion clusters and prominent actin stress fibers at the basal side of the organoids, suggesting the interactions between the cells and the surrounding Matrigel. Finally, our inhibition study showed the dependence of pancreatic ductal organoid rotational migration on myosin activity, actin polymerization, and actin branching.

**Results and discussion**

**Rotation of human pancreatic ductal organoid.**

Normal human pancreatic ductal organoids were derived from the exocrine tissue provided by the Integrated Islet Distribution Program (IIDP). By culturing the exocrine tissue in Matrigel, the ductal cells will grow and establish a spherical structure that consists of a lumen surrounded by a thin layer of epithelial cells (**Figure 1A**)[21,24]. Apicobasal polarity was observed within these cells, where the immunostaining of the apical marker ZO-1 was observed at the lumen side, while the basal marker integrin α6 (ITGA6) was observed between the cells and at the outer side of the cells, which is in contact with the surrounding Matrigel. Moreover, continuous staining of the tight junction protein ZO-1 can be observed between the cells at the apical side (**Figure 1B**). The cystic morphology of the pancreatic ductal organoid is also another evidence of the functional epithelial barrier. Lumen formation is initiated by the secretion of ions (such as the sodium cation and chloride anion) into the luminal space, which increases the luminal osmolality and in turn, drives the influx of water and the increase in lumen volume[25-27]. Hence, the maintenance of this cystic morphology is highly dependent on the epithelial barrier function that preserves the osmotic gradient.

Brightfield live imaging of the pancreatic ductal organoids revealed the rotational motion of the organoids over time (**Video 1**). To investigate the rotation further, the nuclei were stained with a non-toxic, cell-permeable, and highly specific live cell DNA probe, SPY650-DNA, followed by live imaging using spinning disk confocal microscopy. Indeed, we observed the directed motion of the nuclei that resulted from the rotation of the whole organoid (**Figure 1C, Video 2**). We observed two key differences between the observed rotation of our pancreatic ductal organoids, MDCK (canine kidney), and MCF10A (human mammary) spheroids. One, rotation in MDCK and MCF10A spheroids is limited to the early, low cell-number stages (<20 cells)[7,12,13], whereas rotation in the pancreatic ductal organoids can still be observed in large organoids (>300 μm in diameter, **Video 1**) that consist of hundreds of cells. Two, rotation in MDCK cells at later stages that involve more cells (>20 cells) is only observed in a gradient of Matrigel concentrations[8,9] – making it a form of durotaxis – whereas the rotation we observed in the pancreatic organoids is within a homogeneous Matrigel matrix.

As a comparison, we also looked into the pathological pancreatic ductal adenocarcinoma (PDAC) organoids. Similar to their normal counterpart, PDAC organoids also have a cystic morphology and the apicobasal polarity is intact (**Figure S1A**). However, the PDAC cells tend to have cuboid cellular morphology, resulting in a cell layer that is thicker and denser than the normal pancreatic ductal organoid. Moreover, stratified cell layers are occasionally observed within the PDAC organoids. To our surprise, the rotational migration phenotype is lost in the PDAC organoids (**Figure S1B, Video 3**), suggesting the potential importance of rotational migration in maintaining normal tissue homeostasis. This observation is also in line with previous studies in breast cancer[12,13].

**Quantification of human pancreatic ductal organoid rotation.**

Tracking the movement of single cells within large organoids in three dimensions remains a significant challenge. To quantify the organoid rotational migration, live imaging data of the nucleus was processed using a deep learning segmentation method and custom tracking algorithms (see Materials and methods). This image-processing pipeline allows us to reconstruct the nucleus positions over time and derive the displacement trajectory of each nucleus within the rotating organoid (**Figure 2A-B**). Next, the velocity of each nucleus at the organoid surface at a given time point was calculated by dividing the displacement by the time covered between the imaging time points (**Figure 2C**). The average velocity is relatively stable over the imaging period (**Figure 2D**). Looking at the velocity distribution, the velocity ranges between 0 to 35 μm/hr (**Figure 2E**), with the average velocity of around 15-20 μm/hr (**Figure 2F**). This rotation velocity falls within the range of velocity reported in other rotational migration models, including the *Drosophila* follicle cells at 12 to 36 µm/hr[28], the murine pancreatic organoid at around 12 µm/hr[11], the canine kidney epithelial line MDCK at 5 to 20 µm/hr[9], and the human breast epithelial line MCF10A at 12.5 to 17.5 µm/hr[13]. In contrast, the trajectory reconstruction of the PDAC organoids showed that the nuclei are moving randomly within a limited region, at an average velocity of about 3 µm/hr (**Figure S1B-G**).

**Pancreatic ductal organoid rotational migration relies on actomyosin contractility.**

Next, we pondered how the pancreatic ductal cells migrate along the organoid rotational axis. The apical side of the organoid interacts with the fluid-filled luminal space, hence, it is less likely that the cells are migrating on the apical side. On the other hand, the basal side interacts with the Matrigel, and the cells can interact with the ECM proteins through the integrins and migrate on this interface. Indeed, we observed clusters of focal adhesion protein paxillin and prominent actin stress fibers at the basal side of the organoids (**Figure 3A**), indicative of cell-ECM adhesions. Moreover, we also observed the enrichment of the phosphorylated myosin regulatory light polypeptide 9 (pMYL9) on the actin stress fibers (**Figure 3B**), suggesting the activation of non-muscle myosin IIA and actomyosin contractility. This is in line with the findings in the rotating *Drosophila* follicle cells, where paxillin clusters and myosin-decorated actin fibers were also observed at the basal side of the follicle cells[3]. These observations suggest the potential involvement of actomyosin activity in regulating rotational migration.

To investigate the role of myosin contractility, the pancreatic ductal organoids were treated with a small molecule myosin II inhibitor, Blebbistatin that has a high affinity towards the myosin-ADP-Pi complex and locks the myosin in an actin detached state[29]. Nuclei live imaging was performed before and after the Blebbistatin treatment on the same organoid (**Figure 4A, Video 4**). From the trajectory analysis (**Figure 4B**), we observed a significant slowdown of the rotation while the rotational direction was more or less maintained. Averaging the quantified velocity over the imaging period clearly showed a significant decrease in velocity after the Blebbistatin treatment (**Figure 4C-D**). This suggests the reliance of rotational migration on myosin contractility. To note, myosin inhibition does not compromise the luminal structure of the organoid, where the apicobasal polarity was intact and we still observed the continuous tight junction network at the apical side (**Figure S2A-B**). Looking closer at the basal side of the Blebbistatin-treated organoids, we observed significant disruption of the actin stress fibers and the accumulation of pMYL9 at the base of the actin fibers (**Figure 4E**). Instead of the long and relatively straight actin stress fibers observed in the untreated organoids (**Figure 3**), the actin fibers tend to have wavy morphology (**Figure 4E**) and are shorter after the Blebbistatin treatment (**Figure S2C**). These observations suggest that besides myosin activity, actin organization might also be important for organoid rotational migration.

Next, to confirm that rotational migration depends on myosin contractility, we inhibited myosin activity using an orthogonal approach. Organoids were treated with Y-27632, a widely used Rho-kinase (ROCK) inhibitor. ROCK enhances myosin activity by phosphorylating the myosin regulatory light chain (directly activating myosin) and the myosin-binding subunit of myosin phosphatase (suppressing phosphatase activity)[30-32]. Thus, ROCK inhibition reduces myosin phosphorylation, activation, and contractility. Consistent with this, Y-27632 treatment significantly reduced phosphorylated MYL9 levels (**Figure S3A**), and quantification of rotation velocity confirmed a slowdown in migration (**Figure S3B**, **Video 5**). Together, these results confirm that organoid rotational migration depends on myosin activity.

According to the "go-or-grow" hypothesis[33-35], cells favor either proliferation or migration at a given time. Since pancreatic ductal organoid cells proliferate continuously (**Figure S2D**), it is plausible that cell division could impede rotational migration. Indeed, we occasionally observed dividing cells during nuclei live imaging (green arrows in **Video 4**). However, closer inspection of nuclei before Blebbistatin treatment (**Video 4, left**) showed that dividing cells migrated at a speed comparable to neighboring cells, and their division did not noticeably slow rotational movement. This is intriguing, because dividing cells typically round up (**Figure S2E**) and detach from the substrate, suggesting that neighboring cells may help move the dividing cells to maintain rotational continuity. Although less likely, one alternative explanation is that the slowdown observed after Blebbistatin treatment could result from increased proliferation suppressing migration. To test this, we blocked proliferation using Thymidine (**Figure S2D**) prior to Blebbistatin treatment. Even in the absence of proliferation, Blebbistatin still caused a clear reduction in rotation (**Video 6**), indicating that its effect on rotational migration is independent of cell proliferation.

**Rotational migration also depends on actin branching activity.**

The results above pointed towards the critical role of actin organization on pancreatic ductal organoid rotational migration. To study this, we treated the organoids with an actin polymerization inhibitor, Latrunculin A. Indeed, inhibition of actin polymerization significantly slows down the organoid rotation (**Figure S4A**, **Video 7**). Similar to the Blebbistatin-treated organoids, Latrunculin

A treatment does not disrupt the apicobasal polarity of the organoids (**Figure S4B**). However, we observed wavy patterns of the tight junction proteins ZO-1 (**Figure S4C**), which were not observed in the untreated and Blebbistatin-treated organoids. These wavy tight junction structures might imply a decrease in cellular tension at the apical side of the Latrunculin A treated cells. Looking closer to the basal side, Latrunculin A treatment seems to drive the actin distribution towards the perimeter of the cells, causing the enrichment of actin at the cell-cell boundaries (**Figure S4D**). Additionally, the enrichment of activated myosin (pMYL9) was also observed at the periphery of some cells. Combined, these results suggest the importance of actin organization in the organoid rotational migration.

Next, we focused on actin branching, a key process in actin filament assembly and organization. Actin branching is critical for lamellipodium formation—broad, flat protrusions at the leading edge of migrating cells[36]. In rotating Drosophila follicle cells, lamellipodia also form at the leading edge and extend beneath neighboring cells as cryptic lamellipodia[2,37]. By imaging the cell–cell junction protein E-cadherin, we similarly observed cryptic lamellipodia in pancreatic ductal organoids (**Figure 5A**). To test whether actin branching is required for rotational migration, we treated organoids with the Arp2/3 inhibitor CK-666[38]. Since actin branching nucleation depends on assembly of the Arp2/3 complex on existing actin filaments[39], inhibition of this complex blocks branching. CK-666 treatment significantly slowed organoid rotation (**Figure 5B**, **Video 8**). At the basal surface, actin stress fibers remained prominent, some decorated with pMYL9 (**Figure 5C**). However, consistent with previous reports[40], CK-666 disrupted both the structure (**Figure 5D**) and dynamics (**Video 9**) of cryptic lamellipodia. Similarly, the non-rotating PDAC organoids also displayed compromised cryptic lamellipodia (**Figure S1H**) when compared to the rotating pancreatic ductal organoids. Together, these results indicate that rotational migration of pancreatic ductal organoids depends on cryptic lamellipodia and actin branching.

**Conclusion**

In this study, we introduce pancreatic ductal organoids as a new primary human model to investigate epithelial cell rotational migration. Using this system, we quantified rotation velocity and showed that this behavior requires myosin contractility, actin polymerization, and actin branching. To enable these measurements, we developed a 3D single-cell tracking method that is transferable to other systems. The organoid is an experimentally versatile system: we can expand it at a 1:2 ratio biweekly and can genetically modify it with lentiviral or adenoviral transduction, thus enabling rapid mechanistic studies with primary human epithelial cells that better recapitulate normal physiology than immortalized lines. We hope that this model system will contribute to future studies on the unresolved regulatory mechanisms of epithelial cell rotational migration. Key questions include what initiates rotation, what sustains collective migration over time, and which downstream pathways and cellular processes depend on this behavior.

It is also important to consider the biological and experimental context of the pancreatic ductal organoid used in this study. First, the organoids were cultured in Matrigel, a hydrogel derived from mouse sarcoma that contains a complex mixture of ECM proteins, particularly laminin and collagen IV, but is known to exhibit batch-to-batch variability. Given that pancreatic ductal cells interact with the ECM (**Figure 3A**), and that ECM proteins have been shown to regulate pancreatic ductal progenitors[41,42] as well as phenotypes of PDAC organoids[43-45], it is reasonable to predict that rotational migration behavior may vary when organoids are embedded in hydrogels of differing ECM composition, such as collagen I gels[44,45] or polyethylene glycol matrices functionalized with defined ECM ligands[43,46]. Notably, however, rotational migration was consistently observed across multiple batches of Matrigel employed in this study, indicating that this behavior is not contingent upon batch-to-batch variability of Matrigel. Second, the organoid culture media contains growth factors, hormones, and supplements that may influence migration—for example, A83-01 and mNoggin inhibit TGF-β signaling. Because TGF-β is known to regulate cell migration[47], its inhibition in culture could confound our results. However, organoid rotation persisted even after removing the TGF-β inhibitors (**Video 10**), suggesting that this behavior is not solely an artifact of TGF-β suppression. In addition, organoid architecture adds complexity: epithelial cells migrate within a spherical tissue enclosing a pressurized lumen, exposing the apical surface to hydrostatic pressure. Both hydrostatic pressure and surface curvature are known to influence cell behavior[48-50], suggesting that these mechanical cues may modulate rotational migration. Taken together, these considerations position pancreatic ductal organoids as a physiologically relevant and versatile platform for dissecting the biochemical and mechanical regulation of epithelial rotational migration.

**Materials and methods**

**Pancreatic ductal organoid culture**

Normal human pancreatic ductal organoids were derived from pancreatic exocrine tissues isolated from cadaver donors (RRID: SAMN17277513 and SAMN38518868), which were acquired through the Integrated Islet Distribution Program (IIDP). IIDP provided 0.5 mL exocrine tissue, which is equal to ~20 million cells. The exocrine cells were seeded into Matrigel (Corning) to allow for the growth and expansion of pancreatic ductal organoids. The Matrigel culture and expansion of pancreatic ductal organoids were performed following the Tuveson Laboratory Murine and Human Organoid Protocols (http://tuvesonlab.labsites.cshl.edu/wp-content/uploads/sites/49/2018/06/20170523_OrganoidProtocols.pdf), which were kindly compiled based on that lab's their previous studies[21,24]. The human wash media and complete feeding media were used, as suggested by the Tuveson Lab protocol. The wash media consists of Advanced DMEM/F-12 (ThermoFisher), supplemented with 10mM HEPES (ThermoFisher), 1X GlutaMAX supplement (ThermoFisher), 100 µg/mL Primocin (Invivogen), and 0.1% BSA (ThermoFisher). The complete feeding media consists of Advanced DMEM/F-12, supplemented with 10mM HEPES, 1X GlutaMAX supplement, 100 µg/mL Primocin, 1X Wnt3a conditioned media (refer to Tuveson lab protocol), 1X R-spondin1 conditioned media (refer to Tuveson lab protocol), 1X B27 supplement (ThermoFisher), 10mM Nicotinamide (Sigma), 1.25mM N-acetylcysteine (Sigma), 100 ng/mL mNoggin (Peprotech), 50 ng/mL hEGF (Peprotech), 100 ng/mL hFGF (Peprotech), 10nM hGastrin I (Tocris), 500nM A83-01 (Tocris), and 1µM PGE2 (Tocris). Considering the cells doubling time, the organoids were passaged 1:2 every 10 to 14 days. For the passage procedure, the organoid containing Matrigel domes was mechanically disrupted by pipetting the complete media up and down within the well plate. The cell suspension was transferred to a 5 mL protein LoBind tube (Eppendorf) and centrifuged at 200xg, 4°C for 5 minutes. The supernatant was aspirated and the cell pellet was resuspended in ice-cold wash media, followed by centrifugation at 200xg, 4°C for 5 minutes. The supernatant was aspirated and the cell pellet was resuspended in ice-cold Matrigel. The Matrigel cell suspension was pipetted onto a pre-warmed well plate (~37°C) to form the Matrigel domes containing the pancreatic ductal cells, e.g. 50 µL Matrigel dome per well of a 24 well plate. Due to the growth limitation of the human pancreatic ductal organoid, the experiments were done on organoids with passage numbers less than 10. The primary tumor PDAC organoids were acquired from ATCC, as part of the Human Cancer Models Initiative, with the organoid name HCM-CSHL-0092-C25. The culture procedure of the PDAC organoids was very similar to the normal pancreatic ductal organoid culture described above, except for the exclusion of PGE2 in the complete feeding media.

**Inhibitor treatment**

Pancreatic ductal organoids were treated with 200 µM Blebbistatin (Sigma), 80 µM Y-27632 (Tocris), 1 µM Latrunculin A (VWR), or 200 µM CK-666 (MedChemExpress). Treatment concentrations were selected based on titration experiments to determine the lowest dose of each drug that slowdown organoid rotation (**Videos 11–14**). For live imaging, organoids were pre-treated for 1 hour, followed by overnight imaging with the inhibitor within the culture media. Of the tested inhibitors, 80 µM Y-27632 and 1 µM Latrunculin A disrupted organoid morphology after the overnight treatment (**Figure S5**). However, washout restored rotational migration for all treatments, including the recovery of organoid morphology after Y-27632 and Latrunculin A exposure (**Video 15**). This suggests that the chosen concentrations caused only limited toxicity within the overnight treatment window and that they are reversible. For immunofluorescence analysis, organoids were treated for 12 hours before fixation. To block DNA replication and cell proliferation, the organoids were treated with 2 µM Thymidine (Sigma) for 4 days before subsequent treatments.

**Immunofluorescence imaging and immunoblot**

Pancreatic ductal organoids within the Matrigel domes were fixed in 4% formaldehyde (EMS) for 15 min, permeabilized by 0.5% Triton-X (Sigma) for 10 min, blocked by 5% BSA (VWR), and incubated overnight in the primary antibodies against ZO-1 (ThermoFisher #61-7300), ITGA6 (EMD Millipore #MAB1378), phosphorylated MYL9 (pSer19) (ThermoFisher #MA5-15163), MYH9 (Proteintech #11128-1-AP), Paxillin (Abcam #ab32084), ITGB1 (DSHB #AIIB2), or E-cadherin (Cell Signaling #3195S). The samples were incubated in the corresponding Alexa Fluor secondary antibodies (ThermoFisher) for 1.5 hours. DNA was stained with 2 µg/mL DAPI (Sigma) for 15 min. Actin filaments were stained with 1 µg/mL Phalloidin-TRITC (Sigma) for 1 hour. To monitor DNA replication, 20 µM EdU (Abcam) was introduced to the organoid culture for 2 hours prior to fixation, followed by the staining of EdU-labelled cells with 100 mM Tris (pH 8.5, Sigma), 1mM $CuSO_4$ (Sigma), 100 µM Cy5 azide dye (Lumiprobe), and 100 mM ascorbic acid (Sigma) for 30 min at room temperature. Samples were thoroughly washed and the DNA was co-stained with DAPI as described above. Confocal imaging was done using a Zeiss LSM 880 system with a 63X/1.4 NA oil immersion objective.

For the immunoblot, organoids from at least two Matrigel domes were harvested and lysed in Pierce RIPA buffer (ThermoFisher) with the presence of phosphatase inhibitors (Sigma). Proteins extracted were denatured with the presence of NuPAGE LDS buffer (ThermoFisher) at 80°C for 10 minutes, followed by protein separation using the NuPAGE Bis-Tris 4-12% gels (ThermoFisher). Membrane transfer was performed using the iBlot2 system (ThermoFisher). The membrane was incubated overnight at 4°C in the primary antibodies against pMYL9 and β-actin (Santa Cruz, #sc-47778), followed by the corresponding IRDye secondary antibodies (LiCor). The membrane was imaged using the Odyssey Dlx Imager (LiCor).

**Live imaging**

For the live imaging of the nuclei, the pancreatic ductal organoids within the Matrigel domes were stained with 1X SPY650-DNA (Cytoskeleton) for 16 hr prior to the live imaging. For live imaging of the cell membrane, the organoids were stained with 5 µg/mL CellMask Orange (ThermoFisher) for 1 hr prior to the live imaging. The corresponding stains were present throughout the live imaging. Live imaging was performed using an Andor Revolution Spinning Disk Confocal system (Oxford Instruments) with either 20X/0.75 NA or 40X/0.8NA objectives, within an incubator chamber maintained at 37°C and 5% $CO_2$. Z-stack images were taken every 10 or 30 minutes with a z-distance of 1 or 2 µm between each stack. Meanwhile, the brightfield live imaging was performed using an Olympus IX71 with a digital sCMOS camera (Prime 95B, Photometrics) and a 10X/0.3 NA objective, within an incubator chamber maintained at 37°C and 5% $CO_2$.

**Image processing and velocity quantification**

To acquire the position of the nuclei, the confocal z-stack images from each time point was subjected to the tracking and segmentation algorithms TrackMate v.7.14.0[51] and StarDist v.0.9.1[52] within Fiji v.2.15.1[53]. The TrackMate-StarDist analysis provides the x, y, and z locations of the detected nuclei segments, and represents each continuous nucleus segment along the z-stack as a single track. Hence, a track of nucleus segments detected here represents one individual nucleus. Next, by using an in-house MATLAB (MathWorks) algorithm, the mid-z location of each nucleus was determined by recording the location of the nucleus segment with the highest mean intensity along each track. At this point, each nucleus of a given confocal z-stack image was represented by an x-y-z coordinate. Finally, the trajectory tracking of the nuclei over time was done by using the particle tracking algorithm[54], which was adapted into MATLAB. The velocity of each nucleus was calculated from the displacement along the trajectory divided by the time needed to cover the distance. To note, occasionally there are nuclei of non-viable cells within the lumen of the organoids (e.g. **Figure 4A**, **5B**, **S3B**, and **S4A**). As the aim of this analysis is to measure the displacement and velocity of the migrating epithelial cells, the nuclei of these non-viable cells were excluded from the analysis. The 3D plot of nuclei position was done using

scatter3sph v.1.5.0.0 in MATLAB. The MATLAB algorithms can be accessed through https://github.com/IriantoLab/Rotational-migration-in-human-pancreatic-ductal-organoids-depends-on-actin-and-myosin-activity.git

**Statistical analysis**

Statistical analyzes were conducted using Welch's *t*-test between sample pairs. The analyzes were performed using GraphPad Prism v.10. All statistical tests were considered significant if $p < 0.05$. The plots show mean ± SEM.

**Acknowledgment**

The authors would like to thank the Confocal Microscopy Laboratory at FSU College of Medicine for access to the confocal microscopes. The authors would like to thank Dr. Terra Bradley for the careful editing of the manuscript. JI and GX were supported by startup funds from Florida State University, awards from the Florida Department of Health's Bankhead-Coley Cancer Research Program (award number 21B11), and Live Like Bella Pediatric Cancer Research Initiative (award number 23L06). YJW was supported by the American Diabetes Association grant #11-22-JDFPM-03 and the Islet Award Initiative (IIDP-IAI; https://iidp.coh.org/Investigators/Islet-Award-Initiative). Human pancreatic exocrine tissues were provided by the NIDDK-funded Integrated Islet Distribution Program (IIDP) (RRID:SCR_014387) at City of Hope, NIH Grant # U24DK098085. The PDAC organoid line is from the Human Cancer Models Initiative (HCMI), https://ocg.cancer.gov/programs/HCMI. DT and XL were supported by the National Science Foundation (grant no. PHY 2310639), the Collie-Welch Chair through the Welch Foundation (F-0019).

**Figure 1. Rotational migration of the human pancreatic ductal organoid.**

**(A)** Confocal section taken at the middle of a human pancreatic ductal organoid showing the cystic morphology of the organoid, a lumen encapsulated by a layer of pancreatic ductal epithelial cells. The organoids retain their apical-basal polarity as shown by the apical marker protein ZO-1 at the lumen side (inside) and the basal marker protein ITGA6 at the organoid interface with the Matrigel (outside). Bar = 50 µm.

**(B)** Maximum projection of the confocal sections taken from the pancreatic ductal organoid, showing the continuous staining of the tight junction protein ZO-1 between the ductal cells, the distribution of the integrin protein ITGA6 at the basal side of the cells, and the distribution of nuclei within the organoid. Bar = 50 µm.

**(C)** Maximum projection of confocal sections taken during the live imaging of pancreatic ductal organoid nuclei, showing the rotational migration of the organoid. Pseudo-colors (red, yellow, green, and cyan) were assigned to several nuclei to illustrate the location of these nuclei over time. Bar = 50 µm.

**Figure 2. Trajectory and velocity of pancreatic ductal organoid rotation.**

**(A)** Overlay of maximum projections of a rotating pancreatic ductal organoid (organoid #1) taken every 30 minutes, between hours 3 and 7 of the live imaging. To illustrate the rotational migration of the cells, the nuclei were color-coded based on the time point they were imaged. The transparent green arrow indicates the general direction of the rotation. The X and Y position coordinates start from the top left of the image. Bar = 50 µm.

**(B)** Three-dimensional reconstruction of the nuclei trajectories within the rotating organoid #1. **(C)** Vector plot showing the direction and the velocity of nuclei displacement between each time point. The length of the arrows is scaled to the velocity of the nucleus. The vector plot is projected to the XY axis to illustrate the rotational migration of organoid #1.

**(D)** The nuclei average velocity remains relatively stable at ~12-22 µm/hr over the imaging period (mean±SEM, N = 3 organoids from 3 independent experiments, n = 21-48 nuclei per organoid from 7 hours of imaging period, i.e. 15 time points).

**(E)** Histogram plot showing the distribution of the recorded velocity. The velocity ranges between 0 to 35 µm/hr, with peaks at ~15-20 µm/hr (N = 3 organoids from 3 independent experiments, n = 21-48 nuclei per organoid from 7 hours of imaging period, i.e. 15 time points).

**(F)** Average velocity of three independent pancreatic ductal organoids, showing a range between ~15 and ~18 µm/hr (mean±SD, N = 3 organoids from 3 independent experiments, n = 21-48 nuclei per organoid from 7 hours of imaging period, i.e. 15 time points).

**Figure 3. Actin stress fibers, focal adhesion, and phosphorylated myosin at the basal side of the pancreatic ductal organoid.**

**(A) (i)** Confocal maximum projection of the basal surface of pancreatic ductal organoid stained for filamentous actin (F-actin), focal adhesion protein paxillin, integrin β1 (ITGB1), and DNA. Prominent actin stress fibers and clusters of focal adhesion protein, paxillin, were observed at the basal surface. Bar = 50 µm. **(ii-iii)** Zoomed-in images from the yellow dash boxes of **Figure 3A(i)** and intensity profile showing the location of paxillin clusters at the end of the stress fibers. The white arrow indicates the direction of the intensity profile and the white dash box indicates the area used to derive the intensity profile. Bar = 5 µm.

**(B) (i)** Confocal maximum projection of the basal surface of pancreatic ductal organoid stained for F-actin, phosphorylated myosin light chain 9 (pMYL9), non-muscle myosin heavy chain 9 (MYH9), and DNA. Colocalization of F-actin and pMYL9 was observed in some of the stress fibers. Bar = 50 µm. **(ii-iii)** Zoomed-in images from the yellow dash boxes of **Figure 3B(i)** and intensity profile confirm the colocalization of actin stress fibers and pMYL9. The white arrow indicates the direction of the intensity profile and the white dash box indicates the area used to derive the intensity profile. Bar = 15 µm.

**Figure 4. Inhibition of myosin activity disrupts pancreatic ductal organoid rotation and actin stress fibers.**

**(A) (i)** Overlay of maximum projections of a rotating pancreatic ductal organoid (organoid #1) taken before the myosin inhibitor, Blebbistatin, treatment. The images were taken every 30 minutes for 5 hours. The transparent green arrow indicates the general direction of the rotation. Bar = 50 µm. **(ii)** Overlay of maximum projections of the same organoid taken 1 hour after the 200 µM Blebbistatin treatment, showing the significant slowdown of the rotation. The images were taken every 30 minutes for 7 hours. Bar = 50 µm. The nuclei were color-coded based on the time when they were taken. The red arrow points to the nuclei of non-viable cells within the lumen of the organoid, which are excluded from the subsequent nucleus tracking analyzes.

**(B)** The nuclei trajectory of the corresponding pancreatic ductal organoid before **(i)** and after **(ii)** Blebbistatin treatment.

**(C)** Histogram plot showing the velocity distribution of the nuclei before and after the Blebbistatin treatment. The plot shows the decrease in cell velocity after the Blebbistatin treatment on organoid #1 (n = 29-59 nuclei per organoid from 5-7 hours of imaging period, i.e. 11-15 time points).

**(D)** Average velocity of three independent pancreatic ductal organoids before and after the Blebbistatin treatment. The decrease in velocity was observed in all three organoids (mean±SD, *$p < 0.0001$, N = 3 organoids from 3 independent experiments, n = 29-87 nuclei per organoid from 3-7 hours of imaging period, i.e. 7-15 time points).

**(E) (i)** Confocal maximum projection of the basal surface of pancreatic ductal organoid treated with 200 µM Blebbistatin for 12 hr and stained for F-actin, pMYL9, MYH9, and DNA. Bar = 50 µm. **(ii-iii)** Zoomed-in images and intensity profile showing the disrupted actin stress fibers and the accumulation of pMYL9 at the base of the stress fibers. The white arrow indicates the direction of the intensity profile and the white dash box indicates the area used to derive the intensity profile. Bar = 20 µm.

**Figure 5. Inhibition of actin branching also disrupt pancreatic ductal organoid rotation. (A) (i)** Confocal maximum projection of the basal surface of non-treated pancreatic ductal organoid stained for the cell-cell junction protein E-cadherin and DNA. Bar = 50 µm. **(ii)** An orthogonal slice was extracted from the region indicated by the white dash line in Figure 5Ai, showing the presence of cryptic lamellipodium at the basal side of the organoid. Bar = 20 µm. **(B) (i)** Overlay of maximum projections of a rotating pancreatic ductal organoid (organoid #1) taken before the actin branching complex Arp2/3 inhibitor, CK-666, treatment. The images were taken every 30 minutes for 5 hours. The transparent green arrow indicates the general direction of the rotation. Bar = 50 µm. **(ii)** Overlay of maximum projections of the same organoid taken 1 hour after the 200 µM CK-666 treatment, showing the significant slowdown of the rotation. The images were taken every 30 minutes for 7 hours. Bar = 50 µm. The nuclei were color-coded based on the time when they were taken. The red arrow points to the nuclei of non-viable cells within the lumen of the organoid, which are excluded from the subsequent nucleus tracking analyzes. **(iii)** Average velocity of three

independent pancreatic ductal organoids before and after the CK-666 treatment. The decrease in velocity was observed in all three organoids (mean±SD, *$p$ < 0.0001, N = 3 organoids from 3 independent experiments, n = 14-109 nuclei per organoid from 5-7 hours of imaging period, i.e. 11-15 time points).

**(C)** Confocal maximum projection of the basal surface of pancreatic ductal organoid treated with 200 µM CK-666 for 12 hr and stained for F-actin, pMYL9, and DNA. Actin stress fibers were still observed after actin branching inhibition by CK-666, and some of the fibers were decorated with the active pMYL9. Bar = 50 µm.

**(D) (i)** Confocal maximum projection of the basal surface of CK-666 treated pancreatic ductal organoid stained for for the cell-cell junction protein E-cadherin and DNA. Cryptic lamellipodium structures are still observed after inhibition of actin branching, but they are less prominent when compared to the non-treated organoids in Figure 5A. Bar = 50 µm. **(ii)** An orthogonal slice was extracted from the region indicated by the white dash line in Figure 5Di, showing the presence of cryptic lamellipodia at the basal side of the organoid. Bar = 20 µm.

**Figure S1. Human pancreatic ductal adenocarcinoma organoids do not rotate.**

**(A)** Confocal section taken in the middle of a human pancreatic ductal adenocarcinoma (PDAC) organoid showing the cystic morphology of the organoid. These organoids also retain their apical-basal polarity as shown by the apical marker protein ZO-1 at the lumen side (inside) and the basal marker protein ITGA6 at the organoid interface with the Matrigel (outside). Compared to the normal pancreatic ductal organoids, the PDAC cells tend to have a more cuboid cell shape, resulting in a thicker and denser cell layer around the lumen. Stratified cell layers are occasionally observed within the PDAC organoids (yellow arrow). Bar = 50 µm.

**(B)** Overlay of maximum projections of a non-rotating pancreatic ductal adenocarcinoma organoid (PDAC organoid #1) taken every 30 minutes, for 7 hours of live imaging. Bar = 50 µm.

**(C)** Trajectories of the nuclei over time from the corresponding PDAC organoid, showing the local and limited displacement of the nuclei.

**(D)** Vector plot showing the direction and the velocity of nuclei displacement between each time point. The length of the arrows is scaled to the velocity of the nucleus. The vector plot is projected to the XY axis to illustrate the random motion of the nuclei within the PDAC organoid.

**(E)** The velocity of each nucleus at each time point was calculated from the displacement covered between time points. The nuclei average velocity remains relatively stable at ~2.5 µm/hr over the imaging period.

**(F)** Histogram plot showing the distribution of the recorded velocity.

**(G)** The average velocity of three independent PDAC organoids (N = 3 organoids from 3 independent experiments, n = 62-206 nuclei per organoid from 4-7 hours of imaging period, i.e. 9-15 time points).

**(H)** Confocal maximum projection of the basal surface of PDAC organoid stained for the cell-cell junction protein E-cadherin and DNA. Yellow arrows pointing at the cryptic lamellipodium structures at the basal side of the PDAC organoids, which are significantly smaller than the ones observed in the normal pancreatic ductal organoids in Figure 5. Bar = 25 µm.

**Figure S2. Inhibition of myosin does not compromise pancreatic ductal organoid basal-apical polarity, but it significantly disrupts the actin stress fibers.**

**(A)** Confocal section of a pancreatic ductal organoid exposed to 200 µM Blebbistatin for 12 hr, followed by staining of the basal marker, ITGA6, the apical marker, ZO-1, and DNA. The organoids retain their apical-basal polarity as shown by ZO-1 signals at the lumen side (inside) and ITGA6 signals at the organoid interface with the Matrigel (outside). Bar = 50 µm.

**(B)** Maximum projection of the confocal sections taken from the Blebbistatin-treated pancreatic ductal organoid, showing the continuous staining of the tight junction protein ZO-1 between the ductal cells and the distribution of the integrin protein ITGA6 at the basal side of the cells. Bar = 50 µm.

**(C)** The continuous length of actin stress fibers at the basal surface of pancreatic ductal organoids drops significantly after Blebbistatin treatment, as shown by the comparison between Figures 3B and 4E (NT = non-treated, mean±SEM, *$p < 0.0001$).

**(D)** Maximum projection of the confocal sections taken from pancreatic ductal organoids exposed to EdU for two hours prior to fixation and EdU Click-iT reaction. Positive EdU staining in the non-treated organoids indicates the actively proliferating cells within the organoids, which are diminished after four days of 2 µM Thymidine treatment.

**(E)** Confocal section of a pancreatic ductal organoid stained for DNA, centrosome marker CDK5RAP2, alpha tubulin, and F-actin. Yellow arrow showing a round up dividing cell. Bar = 50 µm.

**Figure S3. Inhibition of ROCK disrupt pancreatic ductal organoid rotation.**

**(A)** Immunoblot of cell lysate derived from the non-treated pancreatic ductal organoids and the ones treated with Y-27632, against β-actin and pMYL9. The pMYL9 antibody resulted in the positive band at the expected molecular weight of 20 kDa, and the inhibition of ROCK by Y-27632 resulted in the decrease in pMYL9 level (mean±SEM, *$p < 0.0001$, N = 3 cell lysate from 3 independent experiments).

**(B) (i)** Overlay of maximum projections of a rotating pancreatic ductal organoid (organoid #1) taken before the ROCK kinase inhibitor, Y-27632, treatment. The images were taken every 30 minutes for 5 hours. The transparent green arrow indicates the general direction of the rotation. Bar = 50 µm. **(ii)** Overlay of maximum projections of the same organoid taken 1 hour after the 80 µM Y-27632 treatment, showing the significant slowdown of the rotation. The images were taken every 30 minutes for 5 hours. Bar = 50 µm. The nuclei were color-coded based on the time when they were taken. The red arrow points to the nuclei of non-viable cells within the lumen of the organoid, which are excluded from the subsequent nucleus tracking analyzes. **(iii)** Average velocity of three independent pancreatic ductal organoids before and after the Y-27632 treatment. The decrease in velocity was observed in all three organoids (mean±SD, *$p < 0.0001$, N = 3 organoids from 3 independent experiments, n = 39-212 nuclei per organoid from 5 hours of imaging period, i.e. 11 time points).

**Figure S4. Inhibition of actin polymerization disrupt pancreatic ductal organoid rotation.**

**(A) (i)** Overlay of maximum projections of a rotating pancreatic ductal organoid (organoid #1) taken before the actin polymerization inhibitor, Latrunculin A, treatment. The images were taken every 30 minutes for 3 hours. The transparent green arrow indicates the general direction of the rotation. Bar = 50 µm. **(ii)** Overlay of maximum projections of the same organoid taken 1 hour after the 1 µM Latrunculin A treatment, showing the significant slowdown of the rotation. The images were taken every 30 minutes for 7 hours. Bar = 50 µm. The nuclei were color-coded based on the time when they were taken. The red arrow points to the nuclei of non-viable cells within the lumen of the organoid, which are excluded from the subsequent nucleus tracking analyzes.

**(iii)** Average velocity of three independent pancreatic ductal organoids before and after the Latrunculin A treatment. The decrease in velocity was observed in all three organoids (mean±SD, *p < 0.0001, N = 3 organoids from 3 independent experiments, n = 18-54 nuclei per organoid from 3-7 hours of imaging period, i.e. 7-15 time points).

**(B)** Confocal maximum projection of the basal surface of pancreatic ductal organoid treated with 1 µM Latrunculin A for 12 hr and stained for F-actin, pMYL9, and DNA. Actin polymerization inhibition by Latrunculin A resulted in the total disassembly of actin stress fibers and prominent cortical actin distribution was observed around the perimeter of each cell. Accumulation of pMYL9 was also observed at the perimeter of the cells. Bar = 50 µm.

**(C)** Confocal section of a pancreatic ductal organoid exposed to 1 µM Latrunculin A for 12 hr, followed by staining of the basal marker, ITGA6, the apical marker, ZO-1, and DNA. The inhibition of actin polymerization by Latrunculin A does not disrupt the localization of ITGA6 and ZO-1 at the basal and apical sides, respectively. Bar = 50 µm.

**(D)** Maximum projection of the confocal sections taken from the Latrunculin A-treated pancreatic ductal organoid, showing the continuous staining of the tight junction protein ZO-1 between the ductal cells and the distribution of the integrin protein ITGA6 at the basal side of the cells. Curling of the tight junctions can be observed in some cells. Bar = 50 µm.

**Figure S5. The effect of prolonged drug treatments on organoid morphology.**

Brightfield images showing pancreatic ductal organoid morphology after the treatment with 200 µM Blebbistatin, 80 µM Y-27632, 1 µM Latrunculin A, and 200 µM CK-666. The observation was concluded when the organoid morphology was significantly altered by the treatment, such as with Y-27632 and Latrunculin A treatments.

**Video 1**

Brightfield live imaging reveals the rotation of pancreatic ductal organoids over time.

**Video 2**

Live imaging of the nuclei allows us to better visualize the rotational migration of pancreatic ductal epithelial cells. **(Left)** Maximum projection of confocal sections from an organoid stained with SPY-DNA. The images were taken every 30 minutes. **(Right)** The nuclei were color-coded based on the z position of the nucleus.

**Video 3**

Live imaging of the non-rotating PDAC organoid. **(Left)** Maximum projection of confocal sections from a PDAC organoid stained with SPY-DNA. The images were taken every 30 minutes. **(Right)** The nuclei were color-coded based on the z position of the nucleus.

**Video 4**

Live imaging of pancreatic ductal organoid before **(Left)** and after **(Right)** 200 µM Blebbistatin treatment. The images were taken every 30 minutes. The nuclei of the maximum projections were color-coded based on the z position of the nucleus. Green arrows point to nuclei that were dividing during the live imaging period.

**Video 5**

Live imaging of pancreatic ductal organoid before **(Left)** and after **(Right)** 80 µM Y-27632 treatment. The images were taken every 30 minutes. The nuclei of the maximum projections were color-coded based on the z position of the nucleus.

**Video 6**

Brightfield live imaging of pancreatic ductal organoids after four days of thymidine treatment **(Left)**, followed by 200 µM Blebbistatin treatment **(Right)**. Blebbistatin treatment slows down organoid rotation independent of cell proliferation.

**Video 7**

Live imaging of pancreatic ductal organoid before **(Left)** and after **(Right)** 1 µM Latrunculin treatment. The images were taken every 30 minutes. The nuclei of the maximum projections were color-coded based on the z position of the nucleus.

**Video 8**

Live imaging of pancreatic ductal organoid before **(Left)** and after **(Right)** 200 µM CK-666 treatment. The images were taken every 30 minutes. The nuclei of the maximum projections were color-coded based on the z position of the nucleus.

**Video 9**

Live imaging of the basal layer of pancreatic ductal organoid without **(Left)** and with **(Right)** 200 µM CK-666 treatment, showing the differential cryptic lamellipodia dynamic and structure after CK-666 treatment. The live imaging also shows the polarization of cryptic lamellipodia towards the migration direction. The cell membrane and nuclei were stained by CellMask and SPY-DNA, respectively. Confocal images, taken every 10 minutes, were presented as maximum projections over time.

**Video 10**

Brightfield live imaging of pancreatic ductal organoids cultured in media without the TGF-β inhibitors mNoggin and A83-01.

**Video 11**

Brightfield live imaging of pancreatic ductal organoid before and after Blebbistatin treatment at 25, 50, 100, 200, and 400 µM. Evident slow down of rotational migration was observed at 200 and 400 µM.

**Video 12**

Brightfield live imaging of pancreatic ductal organoid before and after Y-27632 treatment at 5, 10, 20, 40, and 80 µM. Evident slow down of rotational migration was observed at 80 µM.

**Video 13**

Brightfield live imaging of pancreatic ductal organoid before and after Latrunculin A treatment at 0.125, 0.25, 0.5, 1, and 2 µM. Evident slow down of rotational migration was observed at 1 and 2 µM.

**Video 14**

Brightfield live imaging of pancreatic ductal organoid before and after CK-666 treatment at 50, 100, 200, and 400 μM. Evident slow down of rotational migration was observed at 200 and 400 μM.

**Video 15**

Brightfield live imaging of pancreatic ductal organoid before and after the treatment with 200 μM Blebbistatin, 80 μM Y-27632, 1 μM Latrunculin A, and 200 μM CK-666. The drug containing media was then replaced with the normal culture media (W.O. = washout). All organoids recovered and resumed rotation after washout, indicating reversibility of the overnight treatment.

# Figure 1

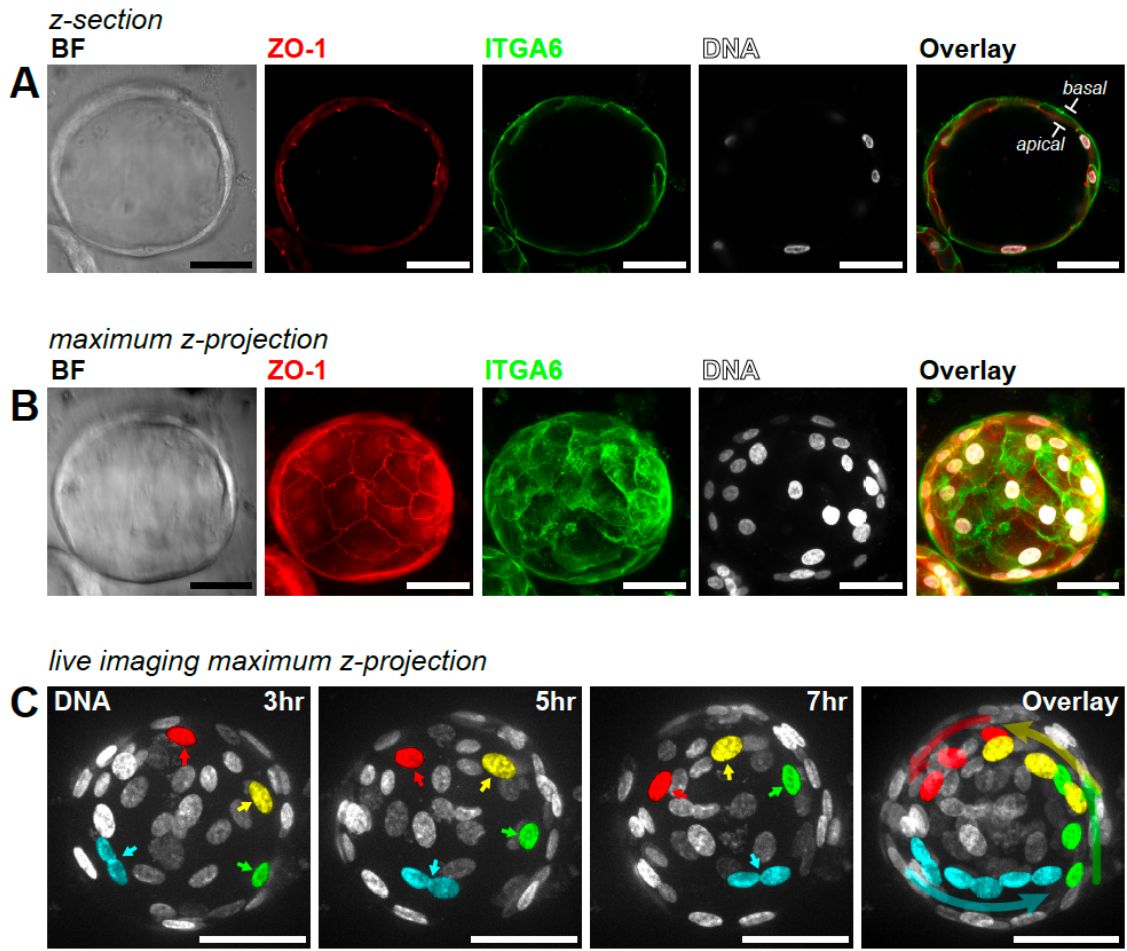

# Figure 2

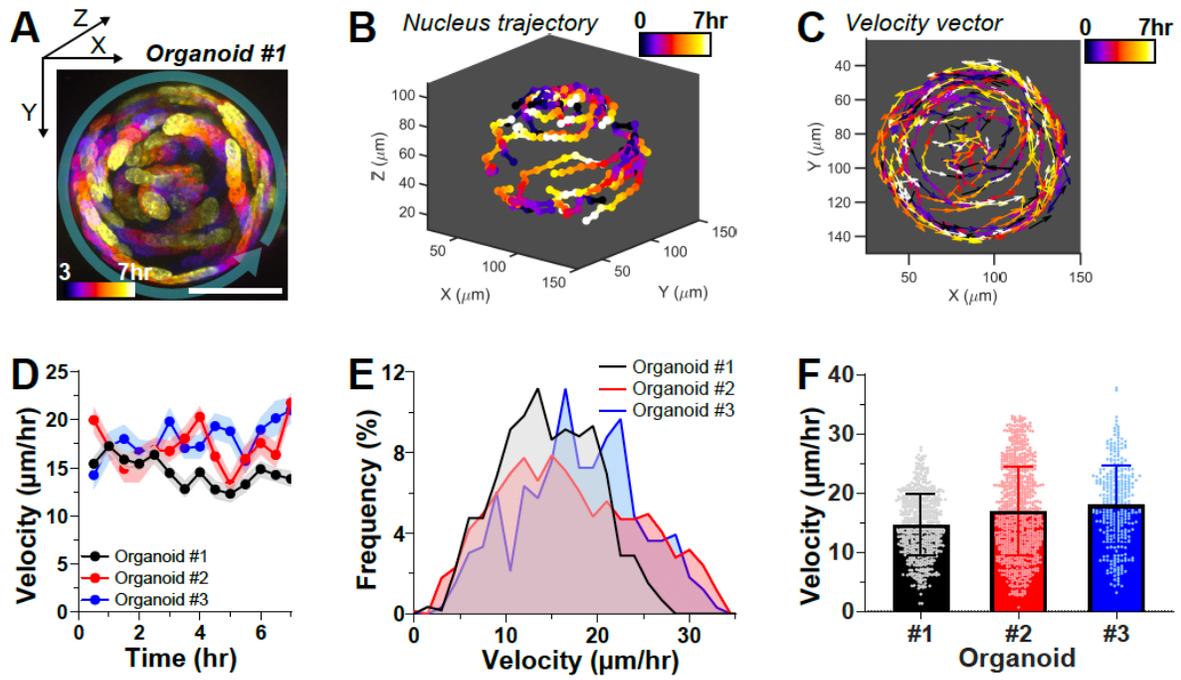

# Figure 3

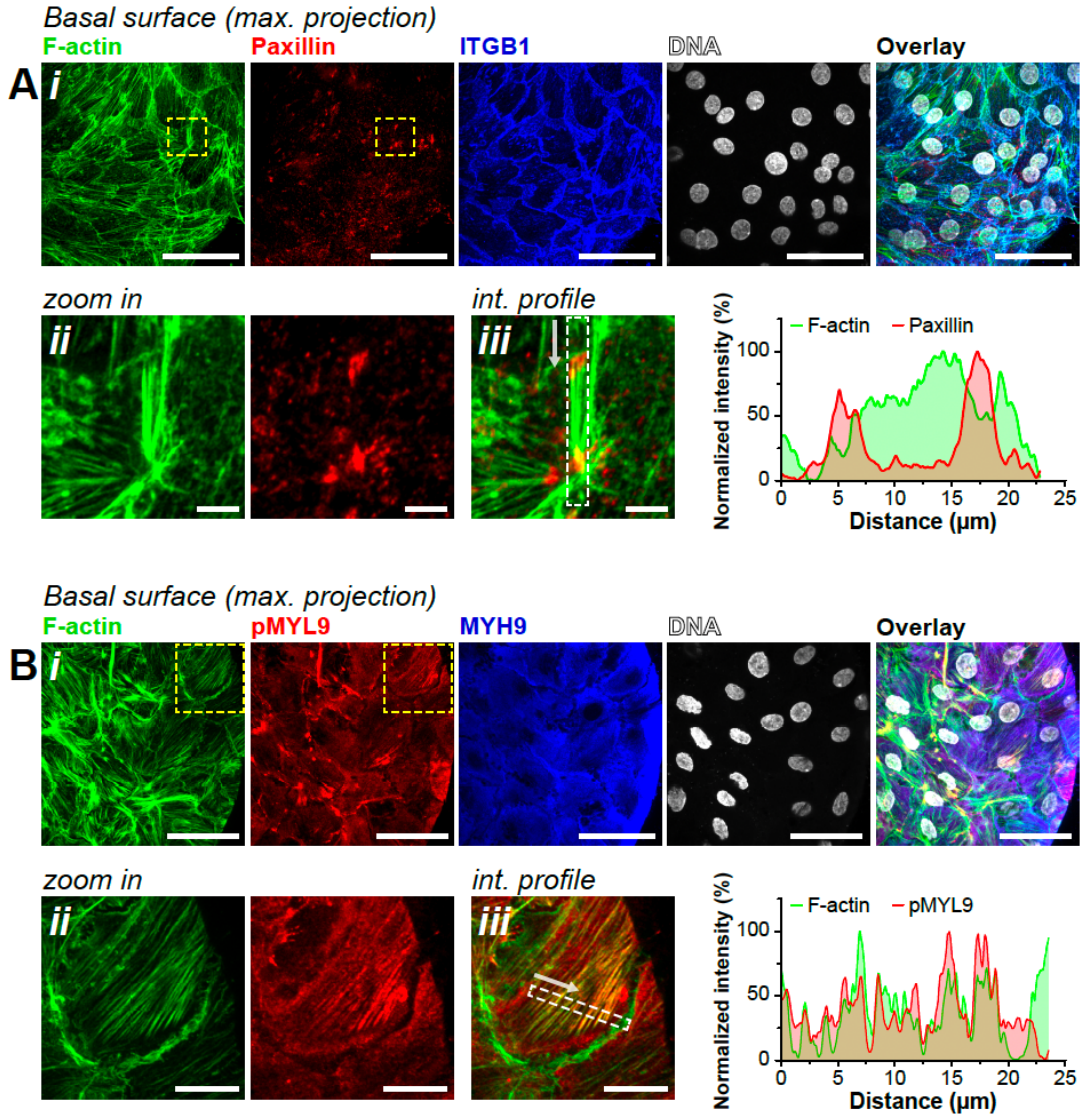

# Figure 4

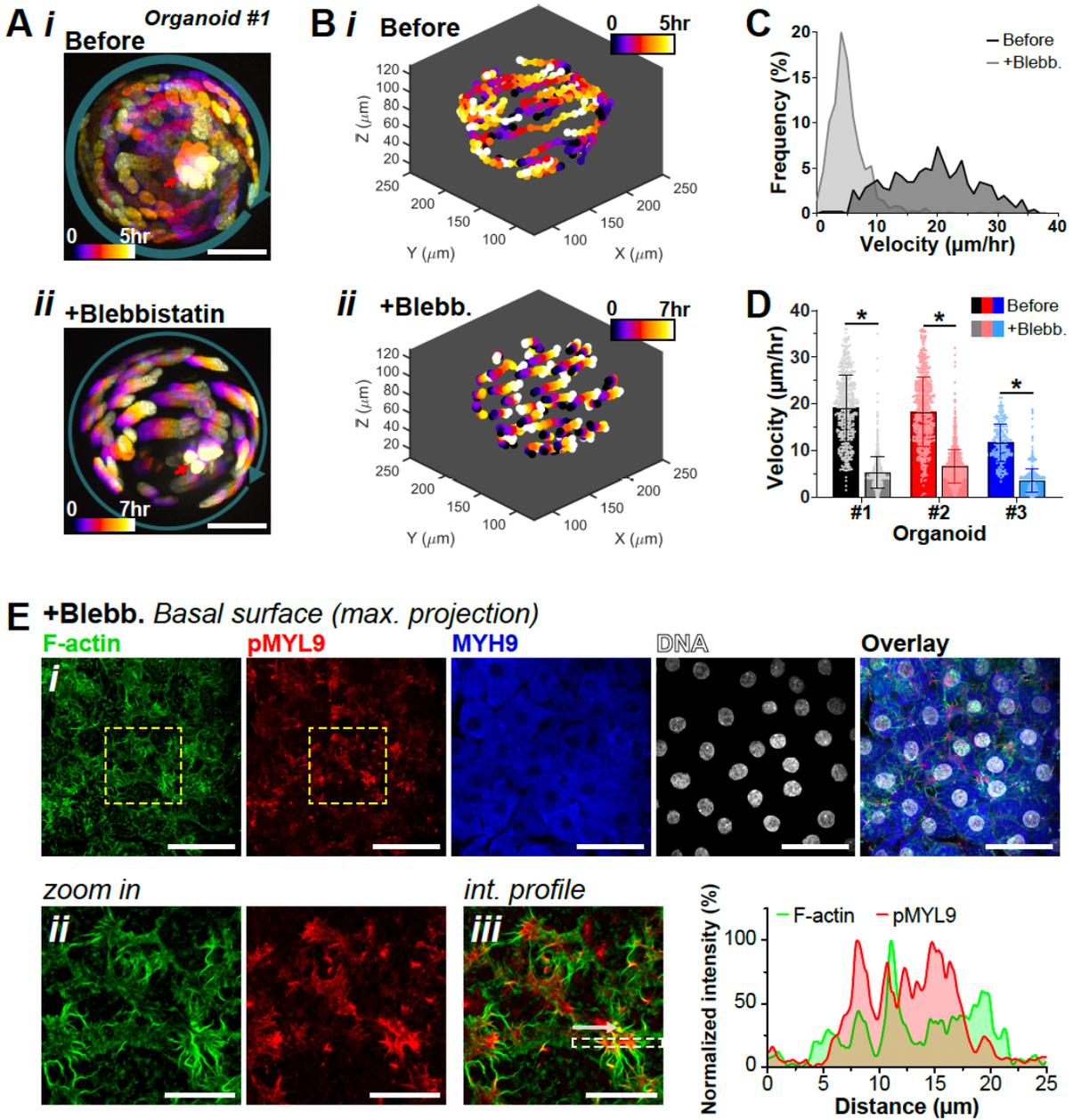

# Figure 5

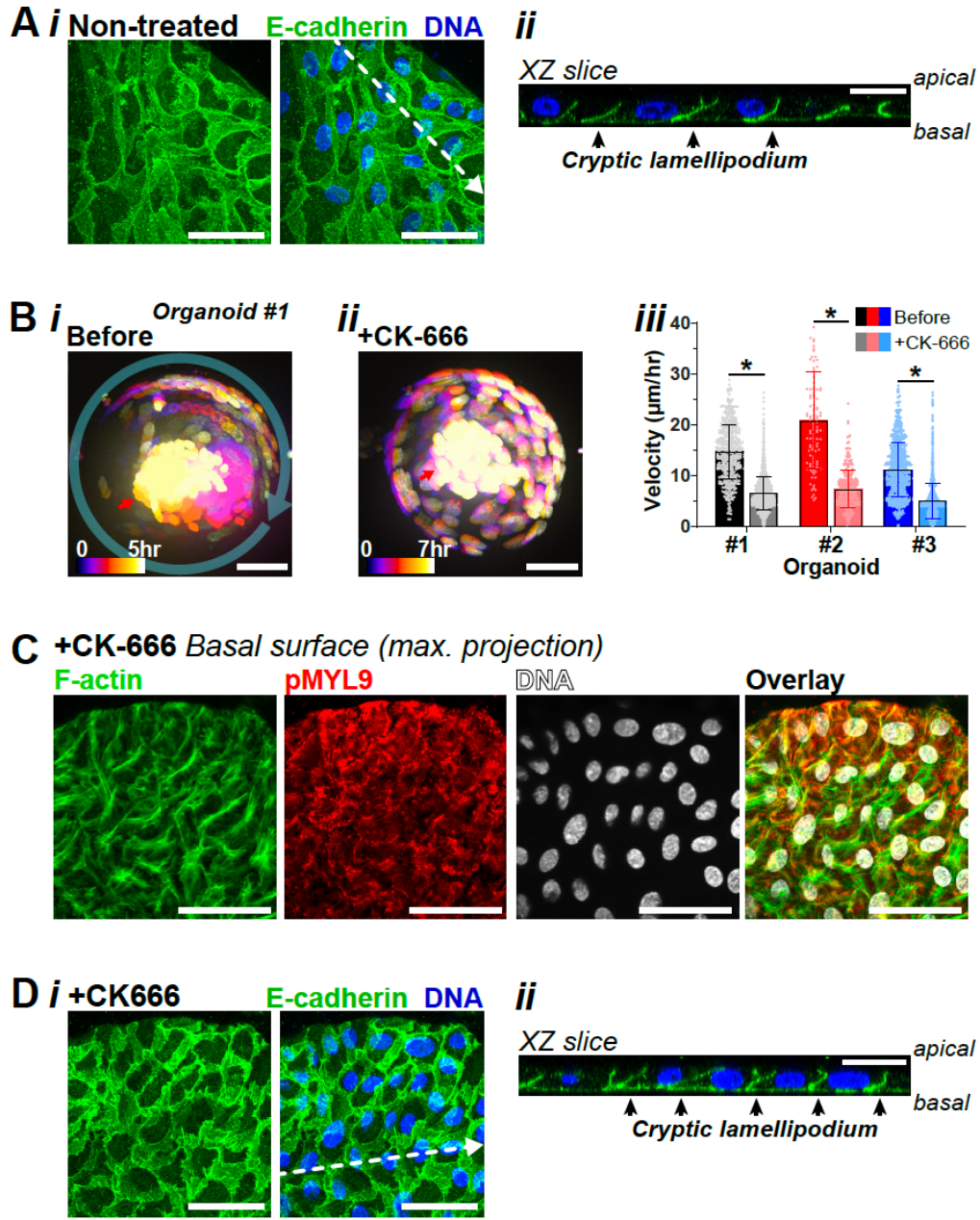

# Figure S1

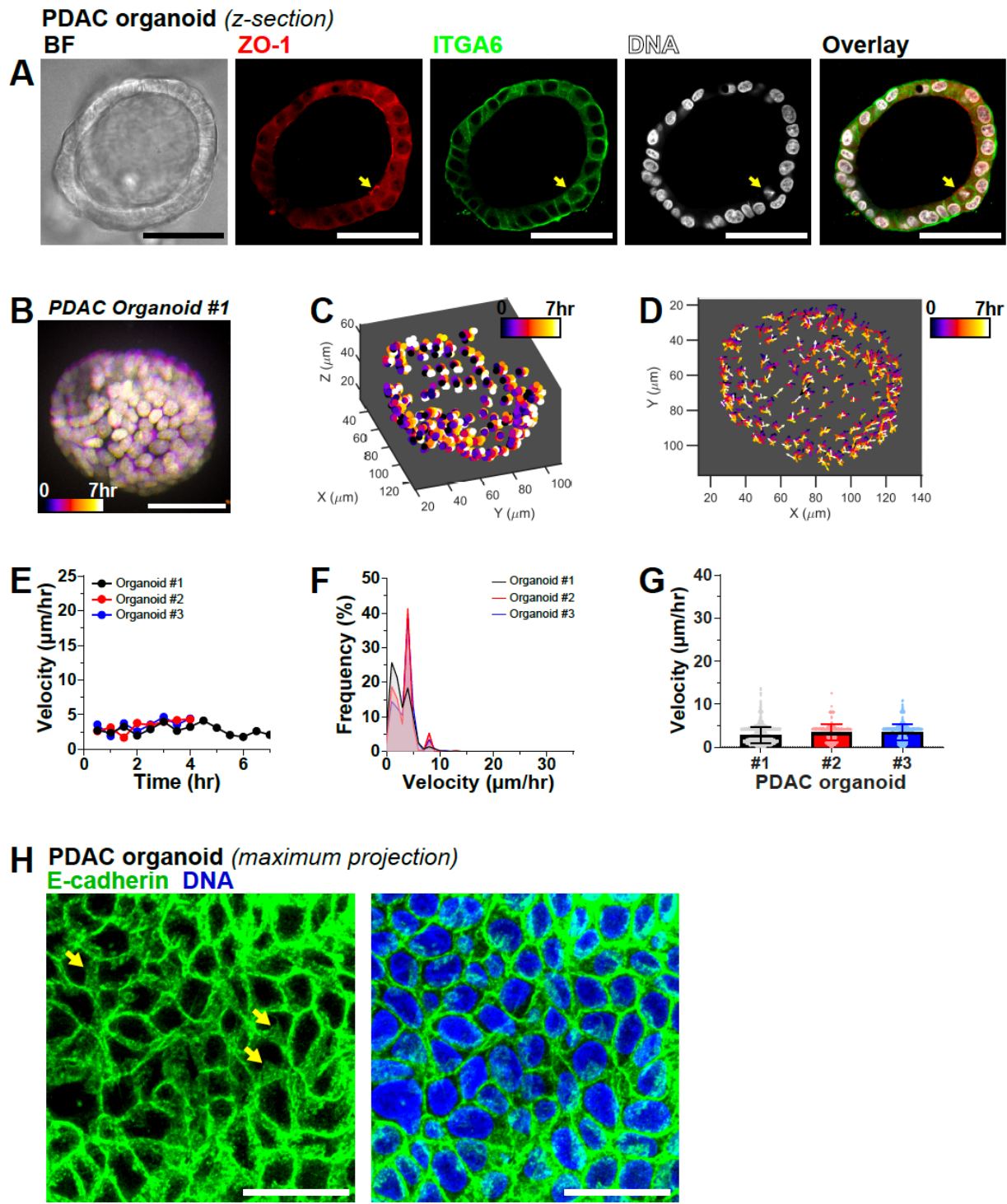

# Figure S2

**A** +Blebbistatin *z-section*
ITGA6　ZO-1　DNA　Overlay

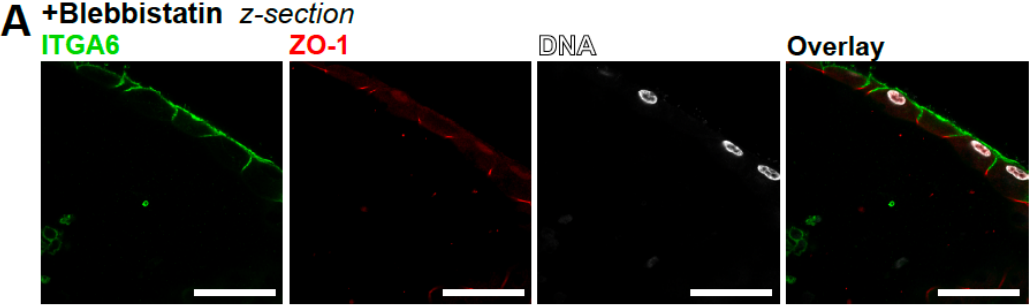

**B** +Blebbistatin *maximum projection*
ITGA6　ZO-1　DNA　Overlay

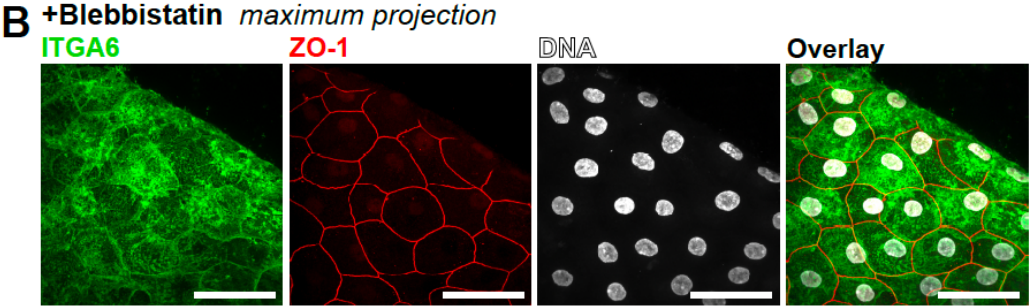

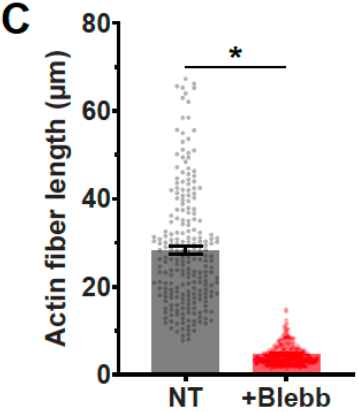

**D** NT *maximum projection*
DNA　EdU　Overlay

4 days Thymidine Block *maximum projection*
DNA　EdU　Overlay

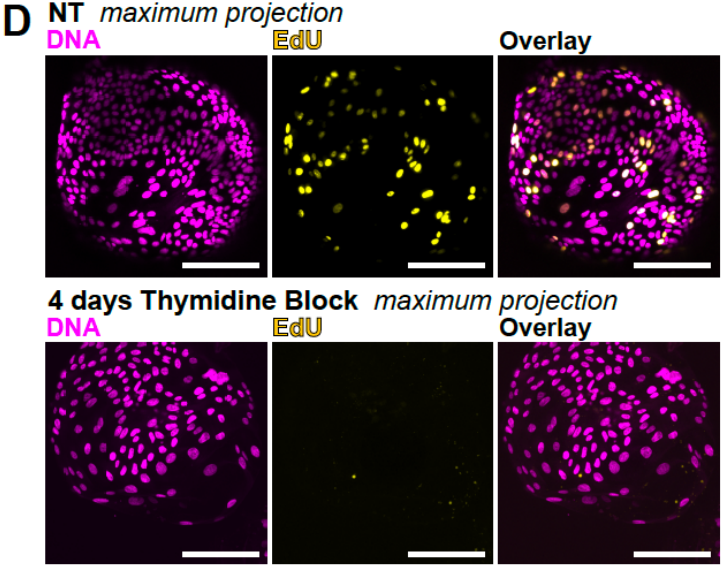

**E** Brightfield　DNA CDK5RAP2 a-tubulin F-actin

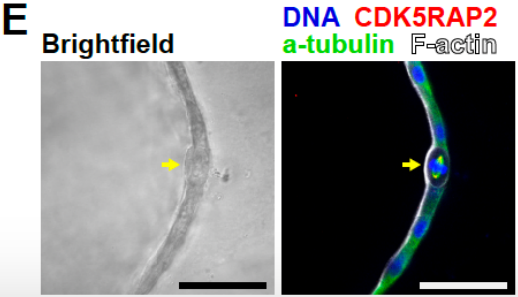

# Figure S3

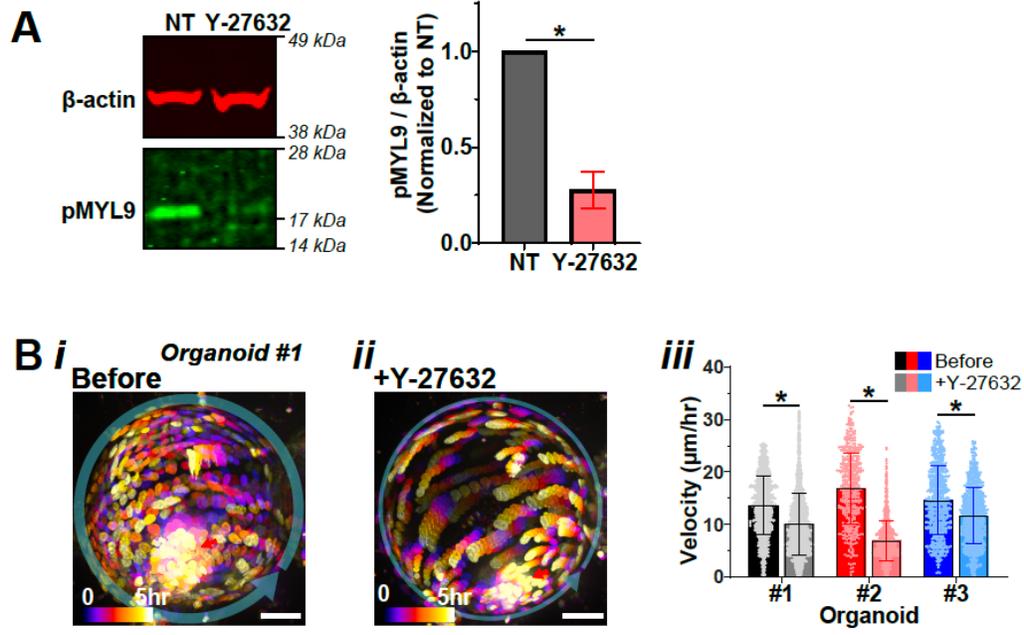

# Figure S4

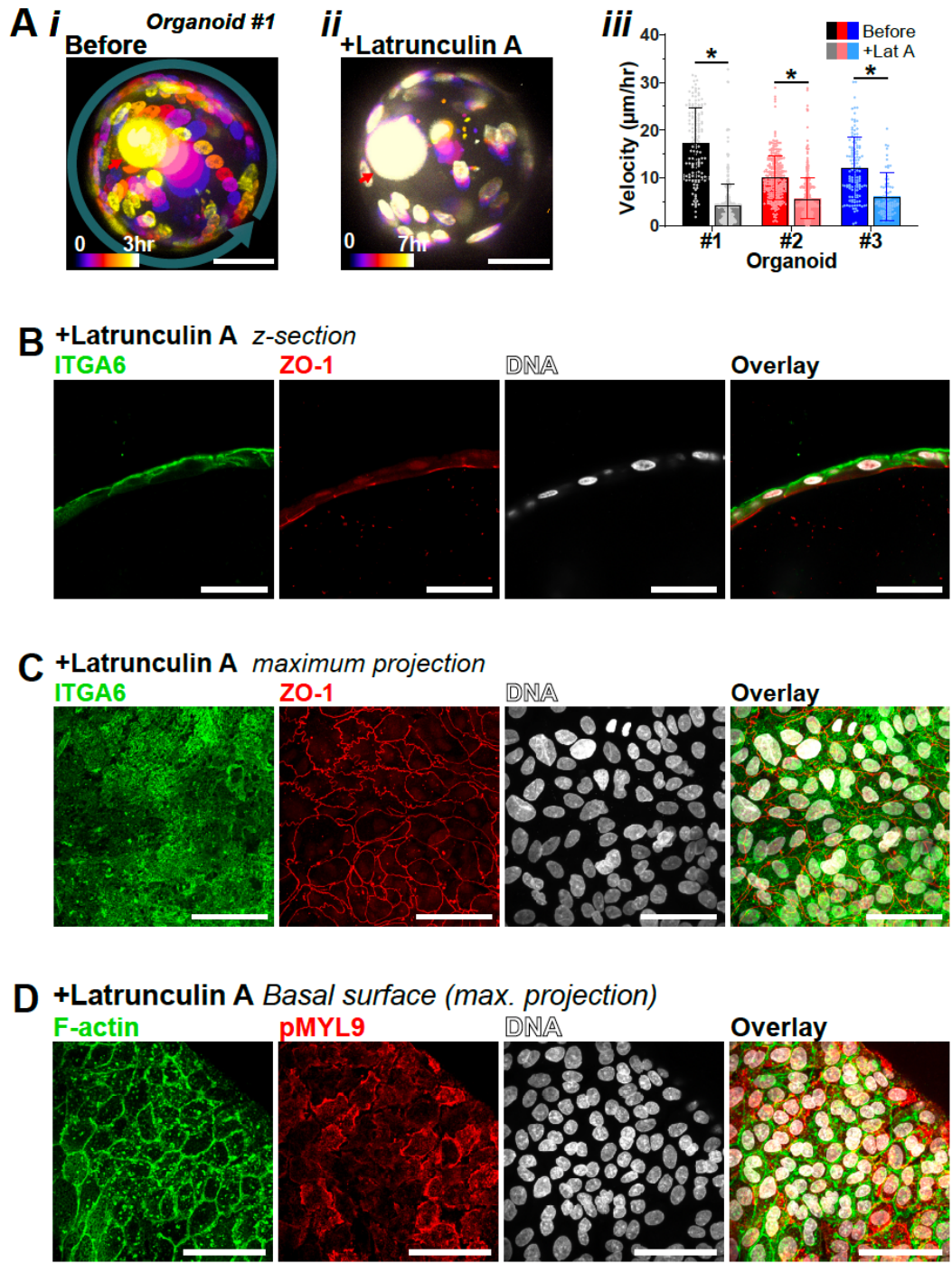

# Figure S5

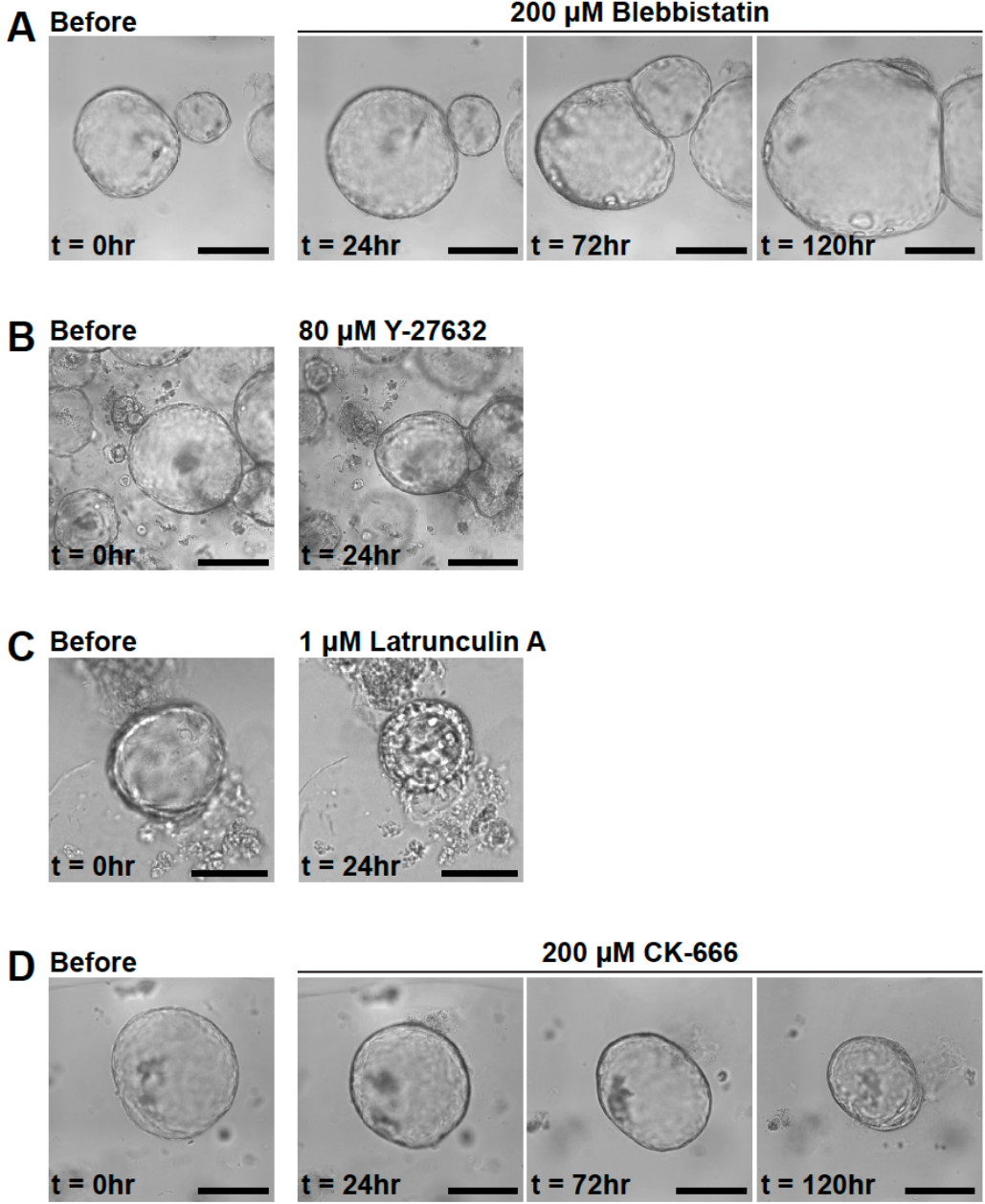

A Before / 200 µM Blebbistatin / t = 0hr, t = 24hr, t = 72hr, t = 120hr

B Before / 80 µM Y-27632 / t = 0hr, t = 24hr

C Before / 1 µM Latrunculin A / t = 0hr, t = 24hr

D Before / 200 µM CK-666 / t = 0hr, t = 24hr, t = 72hr, t = 120hr